\documentclass[preprint,aps,12pt,showpacs,nofootinbib,tightenlines]{revtex4}
\usepackage{amsmath}
\usepackage{amssymb}
\usepackage{epsfig}
\usepackage{graphicx}
\begin{document}
\preprint{NJNU-TH-2005-26}
\newcommand{\beq}{\begin{eqnarray}}
\newcommand{\eeq}{\end{eqnarray}}

\newcommand{\bxsga}{B\to X_s \gamma}
\newcommand{\brbxsga}{{\cal B}(B\to X_s \gamma)}
\newcommand{\bzbzb}{ B_d^0 - \bar{B}_d^0 }

\newcommand{\brbkz}{{\cal B}(B\to \overline{K}^{*0} \gamma)}
\newcommand{\brbkm}{{\cal B}(B\to K^{*-} \gamma)}
\newcommand{\brbrm}{{\cal B}(B\to \rho^- \gamma)}
\newcommand{\brbrz}{{\cal B}(B\to \rho^0 \gamma)}

\newcommand{\calb}{ {\cal B}}
\newcommand{\acp}{ {\cal A}_{CP}}
\newcommand{\oas}{ {\cal O} (\alpha_s)}

\newcommand{\mt}{m_t}
\newcommand{\mw}{M_W}
\newcommand{\mhp}{M_{H}}
\newcommand{\muw}{\mu_W}
\newcommand{\mub}{\mu_b}
\newcommand{\rhob}{\bar{\rho} }
\newcommand{\etab}{\bar{\eta} }

\newcommand{\smallsm}{{\rm \scriptscriptstyle SM}}
\newcommand{\smallnew}{{\rm \scriptscriptstyle new}}
\newcommand{\smallnp}{{\rm \scriptscriptstyle NP}}

\newcommand{\tab}[1]{Table \ref{#1}}
\newcommand{\fig}[1]{Fig.\ref{#1}}
\newcommand{\real}{{\rm Re}\,}
\newcommand{\im}{{\rm Im}\,}
\newcommand{\non}{\nonumber\\ }

\def \epjc{  Eur. Phys. J. C }
\def \jpg{  J. Phys. G }
\def \npb{  Nucl. Phys. B }
\def \plb{  Phys. Lett. B }
\def \prd{  Phys. Rev. D }
\def \prl{  Phys. Rev. Lett.  }
\def \pr{   Phys. Rep. }
\def \rmp{  Rev. Mod. Phys. }
\title{Exclusive $B \to V \gamma$ decays in the T2HDM }
\author{ Zhenjun Xiao}
\email{xiaozhenjun@njnu.edu.cn}
\author{Huihui Cheng}
\author{ and Linxia L\"u }\email{lulinxia@email.njnu.edu.cn}
\affiliation{Department of Physics and Institute of Theoretical Physics,
Nanjing Normal University, Nanjing, Jiangsu 210097,
P.R.China}
\date{\today}
\begin{abstract}
By employing the QCD factorization approach for the exclusive $B
\to V \gamma$ decays, we calculated the new
physics contributions to the branching ratios, CP asymmetries,
isospin and U-spin symmetry breaking of $B \to K^*\gamma$ and $B \to \rho
\gamma$ decays, induced by the charged Higgs penguin diagrams appeared in
the top-quark two-Higgs-doublet model(T2HDM).
Within the considered parameter space, we found that
(a) a charged-Higgs boson with a mass larger than $300$ GeV are always allowed
by the date of $B \to V \gamma$ decay, and such lower limit on $\mhp$ are
comparable with those obtained from the inclusive $B \to X_s \gamma$ decay;
(b) the CP asymmetry of $B \to \rho \gamma$ in the T2HDM can be as large
as $10\%$ in magnitude and has a strong dependence on the angle $\theta$ and the CKM angle
$\gamma$;
(c) the isospin symmetry breakings of $B \to V \gamma$ decays  in the T2HDM
are generally small in size: around $6\%$ for $B \to K^* \gamma$ decay and less than
$20\%$ for $B \to \rho \gamma$ decay;
and (d) the U-spin symmetry breaking $\Delta U(K^*,\rho)$ in the T2HDM is also small
in size, only about $8\%$ of the branching ratio $\calb (B \to \rho^0 \gamma)$.
\end{abstract}

\pacs{13.20.He, 12.60.Fr, 14.40.Nd}

\maketitle

\newpage

\section{introduction}

As is well known, the inclusive radiative decays $B \to X_q
\gamma$ with $q=(d,s)$ and the corresponding exclusive decays $B
\to V \gamma$ ($V=K^*, \rho, \omega $) are very sensitive to the flavor
structure of the standard model (SM) and to the new physics models
beyond the SM \cite{buras96,hurth02,hbb05}.

For the inclusive $B \to X_s \gamma$ decay, the world average of
the experimental measurements \cite{hfag05} is
\beq
{\cal B}(B\to X_s \gamma)^{exp} = (3.39 ^{+0.30}_{-0.27}) \times 10^{-4},
\eeq
which agrees very well with the next-to-leading order (NLO) SM
prediction \cite{hbb05,kn99}. The perfect agreement leads to
strong constraints on many new physics models
\cite{hbb05,carena01,bor00,2hdm,xiao04} where new particles, such
as the charged Higgs boson, charginos  and /or gluinos, may
provide significant contributions to the studied radiative process
through flavor changing loop ( box or penguin ) diagrams.

The exclusive decay mode $B \to K^* \gamma$ has a very clean
experimental signal and a low background, and have been measured
with high accuracy. The world averages of the CP-averaged
branching ratios as given by Heavy Flavor Averaging Group (HFAG)
are \cite{hfag05}
\beq {\cal B}(B^0 \to \overline{K}^{*0} \gamma)
= (40.1 \pm 2.0) \times 10^{-6}, \non
{\cal B}(B^\pm \to K^{*\pm} \gamma)
= (40.3 \pm 2.6) \times 10^{-6}. \label{eq:bvga-exp}
\eeq

The Cabibbo-suppressed $b \to d \gamma$ decay and the
corresponding exclusive $B \to (\rho, \omega) \gamma$ decays were
also measured very recently \cite{hfag05,belle-bd}
\beq
{\cal B}(B\to \rho^+ \gamma) &=& \left ( 0.68 ^{+0.36}_{-0.31}\right )
\times 10^{-6}, \non
{\cal B}(B \to \rho^0 \gamma) &=& \left ( 0.38 \pm 0.18 \right )
\times 10^{-6},\non
{\cal B}(B \to \rho \gamma) &=& \left ( 0.96 \pm 0.23 \right )
\times 10^{-6}, \label{eq:brho-exp} \\
{\cal B}(B \to \omega \gamma) &=& \left ( 0.54 ^{+0.23}_{-0.21}
\right ) \times 10^{-6}. \label{eq:bro-exp}
\eeq
Measurements of these exclusive branching fractions would improve the
constraint on the ratio $|V_{td}/V_{ts}|$ in the context of the
SM, and provide sensitivity to new physics beyond the SM that is
complementary to those from $b \to s\gamma$ and $B \to K^* \gamma$ decays.

When compared with the inclusive $b \to (s,d) \gamma$ decays, the
corresponding exclusive $B \to V \gamma$ decays are experimentally
more tractable but theoretically less clean, since the bound state
effects are essential and need to be described by some
no-perturbative quantities like form factors and light-cone
distribution amplitudes (LCDAs).

In the framework of the SM, the $B \to V \gamma$ decays have been
investigated in leading or next-to-leading order by employing the
constituent quark model (CQM)\cite{deshpande87,greub95}. The
exclusive $B \to (K^*, \rho) \gamma$  decays have been studied by
using a QCD factorization approach
\cite{beneke01,ali02a,bosch02a,bosch02b,kagan02}, or by employing
the perturbative QCD (pQCD) approach \cite{li99,lmsy05}. Such
exclusive decay modes have also been studied recently in some new
physics models beyond the SM \cite{ali01,xz04}.

In this paper,we calculate the new physics contributions to the
branching ratios, CP asymmetries,  the isospin and U-spin symmetry
breaking of the exclusive radiative decays $B \to (K^*, \rho )
\gamma$ in the framework of the top-quark two-Higgs-doublet
model(T2HDM) \cite{das96,wu99,kiers00}. The QCD factorization
method for exclusive $B \to V \gamma$ decays as presented in
Refs.\cite{beneke01,ali02a,bosch02a} will be employed in our
calculations.

This paper is organized as follows. In Sec.~II, we present the
relevant formulas for the calculation of Wilson coefficients, the branching ratio
 ${\calb}(B \to X_s \gamma)$ and some physical observables in the SM and T2HDM.
 In Sec.~III and IV, we calculate the new physics contributions to the $B \to K^*
\gamma$ and $B \to \rho \gamma$ decay in T2HDM, respectively. The
conclusions are included in the final section.

\section{Theoretical framework }\label{sec:th}

For the standard model part, we follow the procedure of
Ref.~\cite{bosch02a} and use the formulas as presented in
Refs.\cite{bosch02a,bosch02b}. The QCD factorization approach to
the exclusive $B \to V \gamma$ decays was applied independently in
Refs.\cite{beneke01,ali02a,bosch02a} with some differences in the
definition and explicit expressions of functions. We adopt the
analytical formulas in the SM as presented in
Refs.\cite{bosch02a,bosch02b} in this paper, since more details
can be found there.

\subsection{Effective Hamiltonian for $b \to s \gamma$}

In the framework of the SM, if we only take into account operators
up to dimension 6 and put $m_s=0$, the effective Hamiltonian for
$b \to s \gamma $ transitions  at the scale $\mu\approx m_b$ reads
\cite{bosch02a} \beq {\cal H}_{eff} =
\frac{G_F}{\sqrt{2}}\sum_{p=u,c}{\lambda_p^s \left
[C_{1}Q_{1}^p+C_{2}Q_{2}^p+\sum_{j=3}^8 C_{j}Q_{j} \right ]}
\label{eq:heff} \eeq where $\lambda_p^q=V_{pq}^{*}V_{pb}$ for
$q=(d,s)$ is the Cabibbo-Kobayashi-Maskawa (CKM) factor
\cite{ckm}. And the current-current, QCD penguin,  electromagnetic
and chromomagnetic dipole operators in the standard basis
\footnote{There is another basis: the CMM basis, introduced by
Chetyrkin, Mosiak, and M\"unz \cite{cmm97} where the fully
anticommuting $\gamma_5$ in dimensional regularization are
employed. The corresponding operators and Wilson coefficients in
the CMM basis are denoted as $P_i$ and $Z_i$ in \cite{bosch02a}.
For the numbering of operators $Q_{1,2}^p$, we use the same
convention as Ref.~\cite{bosch02b} throughout this paper. } are
given by \beq Q_1^p&=&(\bar{s}p)_{V-A}(\bar{p} b)_{V-A}\, , \non
Q_2^p&=&(\bar{s_\alpha}p_\beta)_{V-A}(\bar{p_\beta}
b_\alpha)_{V-A}\, ,\non
Q_3&=&(\bar{s}b)_{V-A}\sum(\bar{q}q)_{V-A}\, ,\non
Q_4&=&(\bar{s_\alpha}b_\beta)_{V-A}\sum(\bar{q_\beta}
q_\alpha)_{V-A}\, , \non
Q_5&=&(\bar{s}b)_{V-A}\sum(\bar{q}q)_{V+A}\, ,\non
Q_6&=&(\bar{s_\alpha}b_\beta)_{V-A}\sum(\bar{q_\beta}q_\alpha)_{V+A}\,
,\non
Q_7&=&\frac{e}{8\pi^2}m_b\bar{s_\alpha}\sigma^{\mu\nu}(1+\gamma_5)b_\alpha
F_{\mu\nu} \, ,\non
Q_8&=&\frac{g}{8\pi^2}m_b\bar{s_\alpha}\sigma^{\mu\nu}(1+\gamma_5)T_{\alpha
\beta}^ab_\beta G_{\mu\nu}^a\,  , \label{eq:qi} \eeq where $T_a$
($a=1,\ldots , 8$) stands for $SU(3)_c$ generators, $\alpha$ and
$\beta$ are color indices, $e$ and $g_s$ are the electromagnetic
and strong coupling constants,
 $Q_1$ and $Q_2$ are current-current operators, $Q_3-Q_6$ are the QCD penguin
operators, $Q_7$ and $Q_8$ are the electromagnetic and
chromomagnetic penguin operators. The effective Hamiltonian for $b
\to d \gamma $ is obtained from Eqs.(\ref{eq:heff}) -
(\ref{eq:qi}) by the replacement $s \to d$.

To calculate the exclusive $B \to V\gamma$ decays complete to
next-to-leading order in QCD and to leading order in
$\Lambda_{QCD}/M_B$, only the NLO Wilson coefficient $C_7(\mub)$
and LO Wilson coefficients $C_i(\mub)$ with $i=(1-6,8)$ and $\mub
= {\cal O}(m_b)$ are needed. For the sake of the readers, we
simply present these Wilson coefficients at the scale $\mu_W =\mw$
and $\mub =m_b$ in Appendix \ref{app:wcs}.

In literature, one usually uses certain linear combinations of the
original $C_i(\mu)$, the so-called ``effective coefficients"
$C^{{\rm eff}}(\mu)$ introduced in Refs.\cite{cmm97,buras94}, in
ones calculation. The corresponding transformations are of the
form \beq
C^{{\rm eff}}_i(\mu) &=& C_i(\mu), \ \ \ (i=1,\ldots , 6), \\
C^{{\rm eff}}_7(\mu) &=& C_7(\mu)   + \sum_{i=1}^6 y_i C_i(\mu),  \\
C^{{\rm eff}}_8(\mu) &=& C_8(\mu)   + \sum_{i=1}^6 z_i C_i(\mu),
\eeq with $\vec{y} = (0,0,0,0,-1/3,-1)$ and
$\vec{z}=(0,0,0,0,1,0)$ in the NDR scheme \cite{buras94}, and
$\vec{y} = (0,0,-1/3,-4/9,-20/3,-80/9)$ and
$\vec{z}=(0,0,1,1/6,20,-10/3)$ in the $\overline{MS}$ scheme with
fully anticommuting $\gamma_5$ \cite{cmm97}. In order to simplify
the notation we will also omit the label ``eff" throughout this
paper.

\subsection{$B \to V \gamma$ decay in the SM}

Based on the effective Hamiltonian for the quark level process $b
\to s(d) \gamma$, one can write down the amplitude for $B \to V
\gamma$ and calculate the (CP-averaged) branching ratios and CP
violating asymmetries once a method is derived for computing the
hadronic matrix elements. By using the QCD factorization approach
\cite{beneke01,ali02a,bosch02a}, one can separate systematically
perturbatively calculable hard scattering kernels ( $T_i^I$ and
$T_i^{II}$ )  from nonperturbative form factors and universal
light-cone distribution amplitudes of $B$, $K^*$ and $\rho$
mesons. The higher order QCD corrections can therefore be taken
into account consistently.

In QCD factorization approach, the hadronic matrix elements of the
operators $Q_i$ with $i=1,\ldots, 8$ for $B \to V \gamma$ decays
can be written as \cite{bosch02a}
\beq
\langle
V\gamma(\epsilon)|Q_i|\bar B\rangle = \left[ F^{B\to V}(0)\,
T^I_{i} + \int^1_0 d\xi\, dv\, T^{II}_i(\xi,v)\, \Phi_B(\xi)\,
\Phi_V(v)\right] \cdot\epsilon \label{eq:bvg1}
\eeq
where $\epsilon$ is the photon polarization 4-vector, $F^{B\to V}$ is
the form factor describing $B \to V$ decays, $\Phi_B$ and $\Phi_V$
are the universal and nonperturbative light-cone distribution
amplitudes for B and $V$ meson respectively \footnote{For explicit
expressions and more details about $\Phi_B$ and $\Phi_V$, one can
see Ref.~\cite{beneke01} and references therein. }, $v$ ( $\bar{v}
\equiv 1-v$ ) is the momentum fraction of a quark (anti-quark)
inside a light meson: $l_1^+ = vk^+$ and $l_2^+ = \bar{v} k^+$
while $k^\mu =(k^+, k^-,\vec{k}_{\bot})$ is a four vector in the
light-cone coordinator, $\xi$ describes the momentum fraction of
the light spectator quark inside a B meson: $l^+ = \xi p_B^+$ with
$\xi = {\cal O}(\Lambda_{QCD}/m_b)$, and $T^I_{i}$ and $T^{II}_i$
denote the perturbative short-distance interactions. The QCD
factorization formula (\ref{eq:bvg1}) holds up to corrections of
relative order $\Lambda_{QCD}/m_b$.

In the heavy quark limit, the contributions to the exclusive $B
\to V \gamma$ decay can be classified into three classes
\cite{bosch02b}: (a) the ``hard vertex" contributions,
(b) the  ``hard spectator"  contributions  and
(c) the ``Weak annihilation" contribution.
Combining these three parts together, the decay amplitude to $\oas$ for
exclusive $B \to V \gamma$ decay takes the form of
\beq
A(\overline{B} \to V \gamma) &=& \frac{G_F}{\sqrt{2}} R_{V}
\langle V\gamma|Q_7|\overline{B} \rangle\, , \label{eq:avga}
\eeq
with
\beq R_{V} =  \lambda_u^{(q)}\,\left [  a_7^u(V\gamma) +
a_{ann}^u (V\gamma) \right ] + \lambda_c^{(q)}\,\left [
a_7^c(V\gamma) +a_{ann}^c (V\gamma) \right ], \label{eq:rv}
\eeq
where $q=s$ for $V=K^*$, $q=d$ for $V=\rho$, and $a_7^p$ ($p=u,c$)
denote the hard vertex and hard spectator NLO contributions
\beq
a^p_7(V\gamma) &=& C^0_7(\mu) + \frac{\alpha_s(\mu) C_F}{4\pi}
\left[ \sum_{i=1,2} Z^0_i(\mu) G_i(z_p)+ \sum_{j=3\ldots 6,8}
Z^0_j(\mu) G_j\right ] \non &&
+ \frac{\alpha_s(\mu_h) C_F}{4\pi}
\left [ C^0_1(\mu_h) H^V_1(z_p) +\sum_{j=3\ldots 6,8} C^0_j(\mu_h)
H^V_j\right ], \label{eq:a7vga}
\eeq
where $z_q = m_q^2/m_b^2$, $\mu_h=\sqrt{0.5 \mu}$, $C_F=4/3$, $Z_j^0(\mu)$ for $j=1\ldots 8$
are the Wilson coefficients defined in the CMM basis \cite{cmm97}.
The explicit expressions of the Wilson coefficients and the
functions $G_i$ and $H^V_j$ can be found in Ref.~\cite{bosch02b}
and in Appendix \ref{app:wcs} and \ref{app:gi}. The functions
$a_{ann}^u$ and $a_{ann}^c$ in Eq.~(\ref{eq:rv}) denote the weak
annihilation contributions and can be found in
Ref.~\cite{bosch02b}.

One special feature of the $B \to \rho \gamma$ decay is that the
weak annihilation can proceed  through the current-current
operator with large Wilson coefficient $C_1$. Although the
annihilation contribution is power-suppressed in $1/m_b$, but it
is compensated by the large Wilson coefficient and the occurrence
of annihilation at tree level.

From the decay  amplitude in Eq.~(\ref{eq:avga}), it is
straightforward to write down the branching ratio for
$\overline{B} \to V \gamma$ decay
\beq
\calb (\overline{B}\to
V\gamma) = \tau_B\frac{G_F^2\alpha m_B^3m_b^2}{32\pi^4} \left
(1-\frac{m_V^2}{m_B^2} \right )^3 \left | R_{V} \right |^2 c_V^2
|F_V|^2,\label{eq:br-vga}
\eeq
where function $R_V$ has been given in Eq.~(\ref{eq:rv}), and $c_V=1$ for $V=K^*, \rho^-$ and
$c_V=1/\sqrt{2}$ for $V=\rho^0$. The branching ratios for the
CP-conjugated $B \to V \gamma$ decay are obtained by the
replacement of $\lambda_p^{(q)} \to \lambda_p^{(q)*}$ in function
$R_V$.

\subsection{Outline of the T2HDM}

Among all the three generation leptons and quarks discovered so
far, the top quark is the unique one: which is much heavier than
all other fermions. Since its discovery in 1995, many
efforts have been made to explain its large mass by considering
the specific Yukawa couplings appeared in the physics models
beyond the SM, the T2HDM \cite{das96,wu99,kiers00} is one of such
kind of new physics models.

The T2HDM is in fact a special case of the third type of
two-Higgs-doublet model, the model III \cite{m3a,m3b,m3x}. In T2HDM, the top
quark is the only fermion receiving its large mass from the vacuum
expectation value(VEV) of the second Higgs doublet $\phi_2$,
$<\phi_2>_{vec}=v_2/\sqrt{2}$ is large. Other five quarks receive
their masses from the VEV of the first Higgs doublet $\phi_1$,
whose VEV $<\phi_1>_{vec}=v_1/\sqrt{2}$ is much smaller.
Furthermore, a new source of CP violation is appeared here:  the
charged Higgs sector of the model contains a CP-violating phase
$``\theta"$ ( $\xi=|\xi|e^{-i \theta}$) in addition to the usual
Cabibbo-Kobayashi-Maskawa(CKM) phase $\delta $ of the SM.

The lagrangian density of Yukawa interactions of the T2HDM can be
simply written as follows \cite{wu99,kiers00}:
\beq {\cal
L}_Y=-\bar{L_L}\phi_1 \, E \, l_R
-\bar{Q_L}\phi_1\, F \, d_R -
\bar{Q_L}\tilde{\phi}_1 \, G\, {\bf 1^{(1)}} u_R
-\bar{Q_L}\tilde{\phi}_2 \, G\, {\bf 1^{(2)}} u_R+H.C.,\label{leff}
\eeq
where the two Higgs doublets are denoted by ${\phi}_{i}$ with $ \tilde{\phi}_i
=i\sigma^2\phi_i^* (i=1,2)$, and where $E, F, G$ are $3\times 3$
matrices in generation space; ${\bf 1}^{(1)}\equiv diag(1,1,0)$,
and $ {\bf 1}^{(2)}\equiv diag(0,0,1)$ are two orthogonal
projection operators onto the first two and the third family
respectively, and $Q_L$ and $L_L$ are the usual left-handed quark
and lepton doublets, while $l_R, u_R$ and $d_R$ are the
right-handed singlets. The heaviness of the top quark arises as a
result of the much large VEV of $\phi_2$ to which no other quark
couples. For this reason, we set that $\tan{\beta}=v_2/v_1 \geq
10$ throughout this paper.

The Yukawa couplings induced by the charged Higgs penguins can be
written as \cite{wu99,kiers00}
\beq
{\cal L}_Y^C&=& \frac{g}{\sqrt{2}\mw}\left\{-\bar{u_L}\,V \, M_D\,  d_R \left[G^+
- \tan{\beta}H^+\right] +\bar{u_R} M_U\, V \,d_L
\left[G^+-\tan{\beta}H^+\right]\right.\non &&+ \left.
\bar{u_R}\Sigma^+ \,V \,d_L
\left[\tan{\beta}+\cot{\beta}\right]H^++H.C.\right\} \
\label{leff2}
\eeq
where $G^{\pm}$ represent the would-be Goldston
bosons, $M_U=diag(m_u,m_c,m_t)$ and $M_D=diag(m_d,m_s,m_b)$ are
the diagonal mass matrices for up- and down-type quarks
respectively, $V$ is the usual CKM matrix and $\Sigma\equiv M_U
U_R^+ {\bf 1^{(2)}} U_R $, and the $U_R^+$ is the unitary matrix which
diagonalizes the right handed up-type quarks.

As defined in Ref.~\cite{wu99}, the matrix $\Sigma$ can be written as
\beq
\Sigma&\equiv &M_U U_R^+ 1^{(2)} U_R  = \left(
\begin{array}{ccc}
0    &    0   &  0\\
0    &    m_c \epsilon_{ct}^2|\xi|^2    &    m_c
\epsilon_{ct}\xi^*\sqrt{1-|\epsilon_{ct}\xi|^2}\\
0    &    m_c\xi^*\sqrt{1-|\epsilon_{ct}\xi|^2}    &
m_t(1-|\epsilon_{ct}\xi|^2) \end{array} \right) \label{eq:sigma}
\eeq
where $\epsilon_{ct} \equiv m_c/m_t$, $\xi=|\xi|
e^{-i\theta}$ is a complex number of order unity. In the framework
of the T2HDM, many studies have been done
\cite{das96,wu99,kiers00}. In this paper, based on previous works
and currently available precision data, we focus on the
calculation of the new physics contributions to the exclusive $B
\to V \gamma$ decays induced by the charge-Higgs penguin diagrams.

\subsection{Wilson coefficients $C^{0}_7$ and $C_8^0$  in the T2HDM}

The new physics contributions to the quark level $b \to (s, d)
\gamma$ transition from the charged Higgs penguins manifest
themselves from the correction to the Wilson coefficients at the
matching scale $\mw$.

For the exclusive decays $B \to V \gamma$ and to the first order
in $\alpha_s$, only the NLO expression for $C_7(\mu)$ has to be
used while the leading order values are sufficient for other
Wilson coefficients appeared in $a_7^p(V\gamma)$ in
Eq.~(\ref{eq:a7vga}). For the SM part, the required Wilson
coefficients can be found in Appendix \ref{app:wcs}. For the T2HDM
part, only the leading order $C_{7}^{\smallnp}$ and $C_{8}^{\smallnp}$ are
known at present \cite{wu99,kiers00} and will be taken into
account in our studies for the exclusive $B \to V \gamma$ decays.

The leading order Wilson coefficients $C_7^{\smallnp}$ and
$C_8^\smallnp$ at the matching scale $\mw$ take the form \cite{wu99,kiers00},
\beq
C_{7}^\smallnp(\mw) & = & \sum_{i=c,t}k^{iq}\left[-\tan^2{\beta}+\frac{1}{m_i
V_{iq}^*}(\Sigma^TV^*)_{iq}\times(\tan^2{\beta}+1)\right]
\left\{B(y_i)+\frac{1}{6}A(y_i)\times \right. \non &&\left.
\left[-1+\frac{1}{m_iV_{ib}}(\Sigma^+V)_{ib}(\cot^2{\beta}+1)\right]\right\},
\label{eq:c70mw-2} \\
C_8^\smallnp(\mw)  & = & \sum_{i=c,t}k^{iq}\left[-\tan^2{\beta}+\frac{1}{m_i
V_{iq}^*}(\Sigma^TV^*)_{iq}\times(\tan^2{\beta}+1)\right]
\left\{E(y_i)+\frac{1}{6}D(y_i)\times\right.\non
&&\left.\left[-1+\frac{1}{m_iV_{ib}}(\Sigma^+V)_{ib}(\cot^2{\beta}+1)\right]\right\},
\label{eq:c80mw-2}
\eeq
where $k^{iq}=-V_{ib}V_{iq}^*/(V_{tb}V_{tq}^*)$, $y_i=(m_i/m_H)^2$, $V$
denotes the CKM matrix, and the matrix $\Sigma$ is defined in
Eq.~(\ref{eq:sigma}). The functions $A$, $B$, $D$ and $E$ in Eqs.(\ref{eq:c70mw-2}) and (\ref{eq:c80mw-2})
are of the form
\beq
A(x) & = &\frac{-7 x + 5 x^2 + 8 x^3}{24(1-x)^3} - \frac{2x^2-3x^3}{4(1-x)^4}\log[x],
\label{eq:aax}\\
B(x) & = &\frac{3x-5x^2}{12(1-x)^2} + \frac{2x-3x^2}{6(1-x)^3}\log[x],
\label{eq:bbx}\\
D(x) & = &\frac{-2x-5x^2 +x^3}{8(1-x)^3} - \frac{3x^2}{4(1-x)^4}\log[x],
\label{eq:c70-xy}\\
E(x) & = &\frac{3x-x^2}{4(1-x)^2}   + \frac{x}{2(1-x)^3}\log[x].
\label{eq:eex}
\eeq

At low energy scale $\mu = {\cal O}(m_b)$, the leading order
Wilson coefficients $C_7^0(\mu)$ and $C_8^0(\mu)$ after the
inclusion of new physics contributions can be written as
\beq
C^0_7(\mu)  & = &  \eta^\frac{16}{23}
 \left [ C_{7,\smallsm}^0(\mw) +C_7^\smallnp(\mw)\right ]\non
&&  +\frac{8}{3} \left (\eta^\frac{14}{23}
-\eta^\frac{16}{23}\right )
                  \left [  C_{8,\smallsm}^0(\mw) + C_8^\smallnp(\mw)\right ]
 + \sum_{i=1}^8 h_i \,\eta^{a_i}\,, \label{eq:c70mb-2}\\
 C^0_8(\mu) & = &   \eta^\frac{14}{23}
 \left [  C_{8,\smallsm}^0(\mw) + C_8^\smallnp(\mw)\right ]
 + \sum_{i=1}^8 \hbar_i\,\eta^{a_i}\, , \label{eq:c80mb-2}
 \eeq
where $\eta=\alpha_s(\mw)/\alpha_s(\mu_b)$, and the ``magic numbers"  $h_i, \hbar_i$ and $a_i$ can be found
in Ref.~\cite{buras96}.

\subsection{ branching ratio ${\cal B}(B \to X_s \gamma$)}

In Refs.\cite{das96,wu99,kiers00}, the authors have
calculated, for example, the new physics corrections to the
electric dipole moment(EDM) of the electron, $F^0-\bar{F}^0$
mixing ($F=K^0, D^0$), $B \to J/\psi K_S$ decay,  and the
branching ratios and CP-violating asymmetries of $b \to (s,d)
\gamma$ decays. Some interesting predictions were found, and the
constraints on the parameter space of the T2HDM were also obtained
by comparing the T2HDM predictions with the data available at that
time \cite{das96,wu99,kiers00}.

The branching ratio of ${B\to X_s \gamma}$ at the NLO level can be written as
\beq
{\cal B}(B \to X_s \gamma)_{NLO} ={\cal B_{SL}}
|\frac{V_{ts}^*V_{tb}}{V_{cb}}|^2 \frac{6\alpha_{em}}{\pi f(z)
k(z)}[|\bar{D}|^2+A+\triangle],
\label{eq:br-sm1}
\eeq
where $\calb_{SL}=(10.64\pm0.23)\%$ is the measured
semileptonic branching ratio of B meson.$\alpha_{em}$=1/137.036 is
the fine-structure pole mass,$z=m_c^{pole}/m_b^{pole}=0.29\pm
0.02$ is the ratio of the quark pole mass,where
$m_b^{pole}=4.8GeV$.The function $f(z)$ and $k(z)$ denote the phase space
factor and the QCD correction for the semileptonic B decay \cite{bg98}.
The term $\bar{D}$ in Eq.~(\ref{eq:br-sm1}) corresponds to the
subprocess  $b\to s\gamma$
\beq
\bar{D}=C_7(\mu_b)+V(\mu_b).
\eeq
The explicit expressions of the function $V(\mu_b)$, $A$ and $\Delta$
can be found easily in Ref.~\cite{bg98}. The term A is the correction coming from the
bremsstrahlung process $b \to s\gamma g$, while the term $\Delta$ includes the non-perturbative
$1/m_b$ and $1/m_c$ corrections. The numerical results show that the new physics contributions
to ``small quantities" $A(\mu_b)$ and $\Delta(\mu_b)$ are very small in magnitude and can be
neglected safely.

Using Eq.~(\ref{eq:br-sm1}) and the input parameters as given in Table \ref{input}, it is easy to calculate
the branching ratio ${\cal B}(B \to X_s \gamma)$. The numerical result is
\beq
{\cal B}(B \to X_s \gamma)_{NLO}^{SM}= (3.52 \pm 0.32) \times 10^{-4}, \label{eq:br-sm}
\eeq
in the SM, and
\beq
{\cal B}(B \to X_s \gamma)^{T2HDM} = (4.00 \pm 0.35 )\times 10^{-4}
\eeq
for fixed $\mhp=400$ GeV, $\xi=1$ and $\tan\beta=30$, here the major errors from different sources have been
added in quadrature.

\begin{figure}[tb]  
\vspace{-1cm}
\centerline{\mbox{\epsfxsize=10cm\epsffile{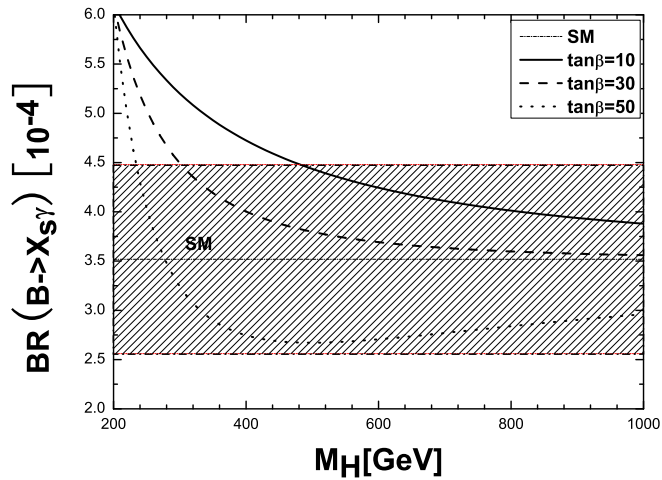}}}
\vspace{-1cm} \caption{The $M_H$ dependence of branching ratio
${\cal B}(B \to X_s \gamma$) in the T2HDM. The gray band shows the
data at $3\sigma$ level. The dot-dashed line refers to the SM
prediction, while the solid, dashed and dots curves show the T2HDM
predictions for $\tan\beta=10,30$ and $50$, respectively. }
\label{fig:fig1}
\end{figure}

In Fig.~\ref{fig:fig1}, we show the $\mhp$ dependence of the
branching ratio in T2HDM directly. The dot-dashed line shows the
central value of the SM prediction, while the solid, dashed and dots curves show the T2HDM prediction for
$\tan\beta=10, 30$ and $50$, respectively.
One can also read out  the lower limits on
$\mhp$ from Fig.~\ref{fig:fig1} directly, for example,
\beq
330 GeV \leq \mhp, \quad for \quad \tan \beta=30.
\eeq
Of course, the lower limit on the mass $\mhp$ has a strong $\tan\beta$ dependence.

\section{$B \to K^* \gamma$ decay }\label{sec:bks}

Now we are ready to calculate the numerical results for the $B \to
V \gamma$ decay in the T2HDM. For the numerical calculations,
unless otherwise specified, we use the central values of the input
parameters as listed in \tab{input}, and consider the
uncertainties of those parameters as given explicitly in
\tab{input}.

\begin{table}
\begin{center}
\caption{Values of the input parameters used in the numerical
calculations \cite{pdg2004,ball98,b03}. For the value of
$F_{K^*}$, we use the lattice QCD determination of $F_{K*} =0.25
\pm 0.06$\cite{b03} instead of the result $F_{K^*}=0.38 \pm 0.06$
as given in Ref.~\cite{ball98}. The smaller value of $F_{K^*}$
gives a better agreement between the SM predictions and the data.
The definition of the Wolfenstein parameter $R_b$ is
$R_b=\sqrt{\bar{\rho}^2+\bar{\eta}^2}$. }
\label{input}
\begin{tabular}{c c c c c c } \\\hline \hline
$ A $    & $\lambda $ & $R_b $         & $\gamma$ &$G_F$ &
$\alpha_{em}$\\ \hline $ 0.854$ & $ 0.2200 $ & $0.39\pm 0.06$ &
$(60\pm 14)^\circ$ & $1.1664\times 10^{-5}GeV^{-2} $ & $ 1/137.036$
\\ \hline \hline
$\alpha_s(M_Z)$& $ m_W  $     & $m_t  $&
$\Lambda^{(5)}_{\overline{MS}}$   & $m_c(m_b) $      &$ m_u $ \\
\hline $0.119$        &$ 80.42$ GeV  & $174.3 $ GeV  & $225$ MeV &
$1.3\pm 0.2$ GeV &  $4.2$ MeV\\ \hline\hline $ f_B    $&
$\lambda_{B} $      & $m_{B_d} $ &$m_b(m_b) $ &$\tau_{B^+}$ &$
\tau_{B^0}$  \\  \hline
$200$ MeV & $(350 \pm 150)$ MeV & $5.279$ GeV&$ 4.2\pm 0.2$ GeV  & $ 1.671 ps$  & $1.536 ps $  \\
\hline \hline
$ F_{K^*}$     & $f_{K^*} $ &$f_{K^*}^{\bot}$ & $m_{K^*}$ &$\alpha_1^{K^*}$ &$\alpha_2^{K^*}$  \\
\hline
$0.25 \pm 0.06$& $230$ MeV  &$185$ MeV        & $894$ MeV &$0.2$ & $0.04$ \\
\hline \hline
$ F_{\rho}$     & $f_{\rho} $ &$f_{\rho}^{\bot}$ & $m_\rho$ &$\alpha_1^{\rho}$ &$\alpha_2^{\rho}$  \\
\hline
$0.29 \pm 0.04$& $200$ MeV  &$160$ MeV        & $770$ MeV &$0$ & $0.2$ \\
 \hline \hline
\end{tabular} \end{center}
\end{table}

From Eqs.(\ref{eq:avga}) and (\ref{eq:br-vga}), the decay
amplitude and branching ratio for $B \to K^* \gamma$ decay can be
written as
\beq
A(\overline{B} \to K^* \gamma) &=&\frac{G_F}{\sqrt{2}} R_{K^*}
\langle K^* \gamma|Q_7|\overline{B} \rangle\, , \label{eq:avks}\\
\calb (\overline{B}\to K^*\gamma) &=& \tau_B\frac{G_F^2\alpha
m_B^3m_b^2}{32\pi^4} \left (1-\frac{m_{K^*}^2}{m_B^2} \right )^3
\left | R_{K^*} \right |^2  \left |F_{K^*} \right |^2,\label{eq:br-ks}
\eeq
with
\beq R_{K^*} =  V_{us}^* V_{ub}\,\left [  a_7^u(K^*\gamma) +
a_{ann}^u (K^*\gamma) \right ] + V_{cs}^* V_{cb}\,\left [
a_7^c(K^* \gamma) +a_{ann}^c (K^*\gamma) \right ].
\label{eq:rks}
\eeq

The CP asymmetry of $B \to K^* \gamma$ can also be defined as
\cite{bosch02b}
\beq \acp(K^* \gamma) =\frac{\Gamma (B \to K^*
\gamma) -\Gamma ( \overline{B} \to \overline{K}^* \gamma)}{
\Gamma( B \to K^* \gamma) + \Gamma(\overline{B} \to \overline{K}^*
\gamma)} \label{eq:acp-vks}
\eeq

Another physical observable for $B \to V \gamma$ decay is the
isospin symmetry breaking in the $K^{*\pm} -\overline{K}^{*0}$ or
$\rho^\pm - \rho^0$ system. Since the branching ratios of both
$B^- \to K^{*-} \gamma$ and $\overline{B}^0 \to \overline{K}^{*0}
\gamma$ decays have been measured, the study of the isospin
breaking in $B \to V \gamma$ decays becomes very interesting now
\cite{ali01,kagan02}. Following Ref.~\cite{kagan02}, the breaking
of isospin symmetry in the $K^{*-} -\overline{K}^{*0}$ system can
be defined as
\beq
\Delta_{0-}(K^*\gamma)\equiv\frac{\eta_{\tau}\brbkz - \brbkm }
{\eta_\tau \brbkz + \brbkm }. \label{eq:iso} \eeq
where $\eta_\tau =\tau_{B^+}/\tau_{B^0}$,
and the CP-averaged branching ratios are understood.

By using the world averages as given in Eq.~(\ref{eq:bvga-exp}) and
the ratio $\tau_{B^+}/\tau_{B^0} = 1.086 \pm 0.017 $
\cite{pdg2004}, we find numerically that
\beq
\Delta_{0-}(K^* \gamma)^{exp} = (3.9\pm 4.2) \% ,
\label{eq:d0m-exp}
\eeq
where the errors from the two measured branching ratios and the ratio
$\tau_{B^+}/\tau_{B^0} $ have been added in quadrature. The
measured value of isospin symmetry breaking is indeed small as
expected previously. Any new physics contribution producing large
isospin breaking for $B \to K^* \gamma $ decays will be strongly
constrained by this measurement.

\subsection{Branching ratios and CP asymmetries}

By using the formulas  as given in Eqs.(\ref{eq:a7vga}) and the
central values of input parameters in \tab{input}, the NLO SM
predictions for branching ratio $\calb (B \to K^*\gamma)$ are
\beq
\calb (B \to \bar{K}^{*0}\gamma )^{SM} &=&
\left [
3.36^{+1.62}_{-1.30}(F_{K^*}) ^{+0.62}_{-0.60} (\mu)
^{+0.23}_{-0.09}(\lambda_B)\pm 0.20 (m_c)\right ] \times 10 ^{-5} \non
&=& \left ( 3.36 ^{+1.76}_{-1.45}  \right ) \times 10^{-5}, \label{eq:brk0sm-s}\\
\calb (B \to K^{*-}\gamma )^{SM} &=&
\left [
3.34^{+1.66}_{-1.32}(F_{K^*}) ^{+0.28}_{-0.47} (\mu)
^{+0.33}_{-0.12}(\lambda_B)\pm 0.20 (m_c)\right ] \times 10 ^{-5}\non
&=& \left ( 3.34 ^{+1.72}_{-1.42}  \right ) \times
10^{-5}, \label{eq:brkpsm-s}
\eeq
where the errors from the uncertainties of the input parameters have been added in
quadrature, the largest theoretical error comes from the uncertainty of the
form factor $F_{K^*}$: $F_{K^*}=0.25 \pm 0.06$. By comparing these theoretical
predictions with the measured values as given in
Eq.~(\ref{eq:bvga-exp}), we can find that the central values are
smaller than the world average, but they are in good agreement
within one standard deviation.

According to previous studies in Ref.~\cite{kiers00,wu99}, we got to know that the charged Higgs
penguins can provide a significant contribution to the dominant
Wilson coefficient $C_7(\mu)$. After taking into account the constraints from the data of
$B \to X_s \gamma$,  a charged Higgs boson with a mass Of  $300-500$ GeV is
still allowed,  as illustrated in Fig.~\ref{fig:fig1}. We now calculate the branching ratios
and CP-violating asymmetries for the $B \to K^* \gamma$ in the T2HDM.

Using the input parameters as given in Table \ref{input}, the theoretical
predictions in the T2HDM are:
\beq
\calb (B \to \overline{K}^{*0}\gamma )^{\rm T2HDM} &=& \left
[ 3.85^{+1.86}_{-1.49}(F_{K^*}) ^{+0.15}_{-0.34} (\mu)
^{+0.26}_{-0.10}(\lambda_B)^{+0.21}_{-0.20} (m_c)\right ] \times
10 ^{-5} \non
&=& \left ( 3.85 ^{+1.90}_{-1.54}  \right ) \times 10^{-5},
\label{eq:brk0-m3a}\\
\calb (B \to K^{*-}\gamma )^{\rm T2HDM} &=& \left [3.77^{+1.92}_{-1.52}(F_{K^*}) ^{-0.11}_{-0.20} (\mu)
^{+0.35}_{-0.14}(\lambda_B)\pm 0.21 (m_c)\right ] \times 10 ^{-5}\non
&=& \left ( 3.77 ^{+1.97}_{-1.55}  \right ) \times 10^{-5},
\label{eq:brkp-m3a}
\eeq
for $\mhp =400$ GeV, $\xi=1$ and $tan{\beta}=30$.

\begin{figure}[htb]  
\vspace{-1cm}
\centerline{\mbox{\epsfxsize=10cm\epsffile{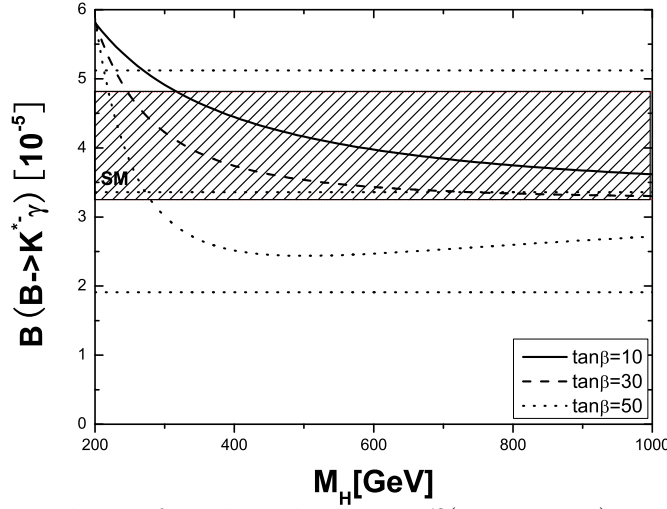}}}
\vspace{-1cm}
\caption{ The $\mhp$ dependence of the branching
ratio ${\cal B}(B \to K^{*-} \gamma)$ in the T2HDM for
$\tan{\beta}=10,30,50$, respectively. More details see the text.}
\label{fig:fig2}
\end{figure}

\begin{figure}[htb]  
\vspace{-1cm}
\centerline{\mbox{\epsfxsize=10cm\epsffile{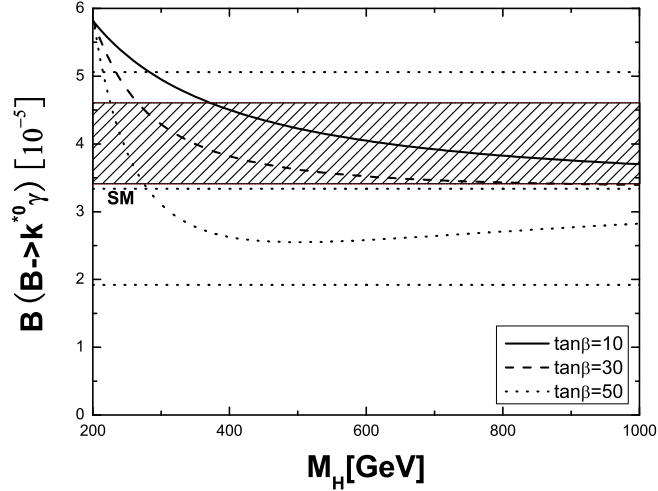}}}
\vspace{-1cm} \caption{ The same as Fig.\ref{fig:fig2}, but for $B
\to \bar{K}^{*0} \gamma$ decay.}
\label{fig:fig3}
\end{figure}

Fig.~\ref{fig:fig2} shows the $\mhp$ dependence of the branching
ratio of $B \to K^{*-} \gamma$ decay in the SM and T2HDM for
$\tan{\beta}=10$ (solid curve),$30$ (dashed curve) and $50$ (dots curve), respectively.
Three horizontal dots lines shows the central value and the $1\sigma$ lower and upper bounds of
the SM prediction: $\calb ( B \to \overline{K}^{*0}
\gamma)=(3.36^{+1.76}_{-1.45})\times 10^{-5}$.
The shaded band shows the region allowed by the data at $3\sigma$ level:
$ 3.25\times 10^{-5} \leq \calb ( B \to K^{*-} \gamma) \leq 4.81\times 10^{-5}$.

Fig.~\ref{fig:fig3} shows the $\mhp$ dependence of the branching
ratio $B \to K^{*0} \gamma$ in the SM and T2HDM for
$\tan{\beta}=10,30,50$, respectively.
Three horizontal dots lines shows the central value and the $1\sigma$ lower and upper bounds of
the SM prediction: $\calb ( B \to \overline{K}^{*0}
\gamma)=(3.34^{+1.72}_{-1.42})\times 10^{-5}$.
The shaded band shows the region allowed by the data at $3\sigma$ level:
$ 3.41\times 10^{-5} \leq \calb ( B \to \bar{K}^{*0} \gamma) \leq 4.61\times 10^{-5}$.
Other curves have the same meaning as in Fig.~\ref{fig:fig2}.

\begin{figure}[htb]  
\vspace{-1cm}
\centerline{\mbox{\epsfxsize=10cm\epsffile{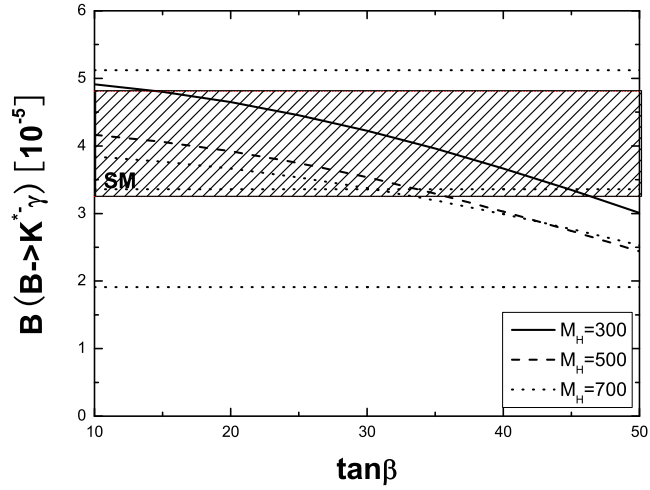}}}
\vspace{-1cm} \caption{ The $\tan{\beta}$ dependence of the
branching ratio $\calb (B \to K^{*-} \gamma)$ in the T2HDM. The
solid, dashed  and dots curve shows the central value of the T2HDM
prediction for $\mhp=300,500,700$ GeV,respectively. }
\label{fig:fig4}
\end{figure}

Fig.~\ref{fig:fig4} shows explicitly the $\tan{\beta}$ dependence of the
branching ratio $B \to K^{*-} \gamma$ in the T2HDM for
$\mhp=300,500,700$ GeV, respectively.
The solid, dashed and dots curve shows the central value of the T2HDM prediction for
$\mhp=300, 500$ and $700$ GeV, respectively.
Other curves in this figure have the same meaning as the corresponding curves in Fig.~\ref{fig:fig2}.
For $B \to \overline{K}^{*0} \gamma$ decay channel, we found a very similar $\tan{\beta}$ dependence.

One can see from Figs.(\ref{fig:fig2}-\ref{fig:fig4}) that a light charged-Higgs boson with a mass less than
$200$ GeV is excluded by the data of $\calb (B \to K^{*0,*-} \gamma)$, while a charged-Higgs boson
with a mass larger than $300$ GeV in the T2HDM is still allowed by the same data.
This lower bound  is well consistent with the one obtained from the data of the inclusive decay
$B \to X_s \gamma$.

\begin{figure}[htb]  
\vspace{-1cm}
\centerline{\mbox{\epsfxsize=10cm\epsffile{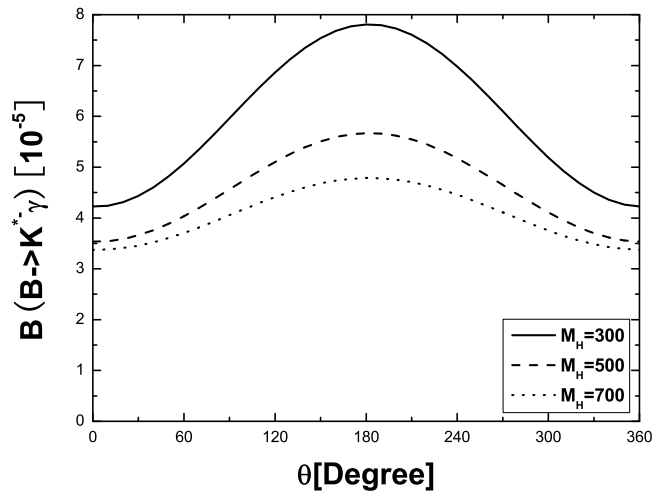}}}
\vspace{-1cm}
\caption{ The $\theta$ dependence of the branching ratio $\calb (B \to K^{*-}\gamma)$
in the T2HDM. For details see the text.} \label{fig:fig5}
\end{figure}

In Fig.~\ref{fig:fig5}, we show the $\theta$ dependence of the branching ratio $\calb (B \to K^{*-}\gamma)$
in the T2HDM for $\tan\beta=30$, $|\xi|=1$, and $\mhp=300$ (solid curve), $500$ (dashed curve) and
$700$ GeV (dotted curve), respectively.

For the exclusive $B \to K^* \gamma$ decay, the theoretical
prediction for the CP-violating asymmetry $\acp$ as defined in
Eq.~(\ref{eq:acp-vks}) is very small:
\beq
\left |\acp(B \to K^{*} \gamma ) \right |  < 1\%
\eeq
in both the SM and the T2HDM considered here.

\subsection{Isospin Symmetry}

In last subsection, one can see that it is the uncertainty of the form factor
$F_{K^*}$ which produces the dominant error for the theoretical
predictions of the branching ratios, but such strong dependence on the form factor $F_{K^*}$
will be largely canceled in the ratio
for the isospin symmetry breaking of $B \to K^* \gamma$ system.
From Eqs.(\ref{eq:rv},\ref{eq:br-vga}), the isospin symmetry breaking
$\Delta_{0-}(K^*\gamma)$ as defined in Eq.~(\ref{eq:iso})
can also be written as
\beq
\Delta_{0-}(K^* \gamma) = \frac{ \left |R_{\overline{K}^{*0}} \right |^2- \left |R_{K^{*-}} \right |^2 }{
\left |R_{\overline{K}^{*0}} \right |^2 + \left |R_{K^{*-}} \right |^2  }
\label{eq:isob}
\eeq
where the function $R_{K^*}$ have been defined in Eq.~(\ref{eq:rks}).

In the SM, it is easy to find the numerical result for $\Delta_{0-}(K^* \gamma)$,
\beq
\Delta_{0-}(K^* \gamma)^{\smallsm} &=& \left [5.8^{+4.1}_{-2.1} (\mu) ^{+1.7}_{-1.0}(F_{K^*})
^{+0.6}_{-1.3}(\lambda_B)^{+0.2}_{-0.1} (m_c)\right ] \times 10^{-2} \non
&=& \left ( 5.8 ^{+4.5}_{-2.7}  \right ) \times
10^{-2}, \label{eq:d0m-sm}
\eeq
where individual  errors have been added in quadrature, and the remaining $F_{K^*}$ dependence comes from
the annihilation contributions which also have a relatively weak $F_{K^*}$ dependence.
The large theoretical error is dominated by the uncertainty of the low energy scale $m_b/2 \leq \mu \leq 2
m_b$. The SM prediction are well consistent with the measured
value of $\Delta_{0-}^{\rm exp}(K^* \gamma) = (3.9\pm 4.2)\% $.

\begin{figure}[htb]  
\vspace{-1cm}
\centerline{\mbox{\epsfxsize=10cm\epsffile{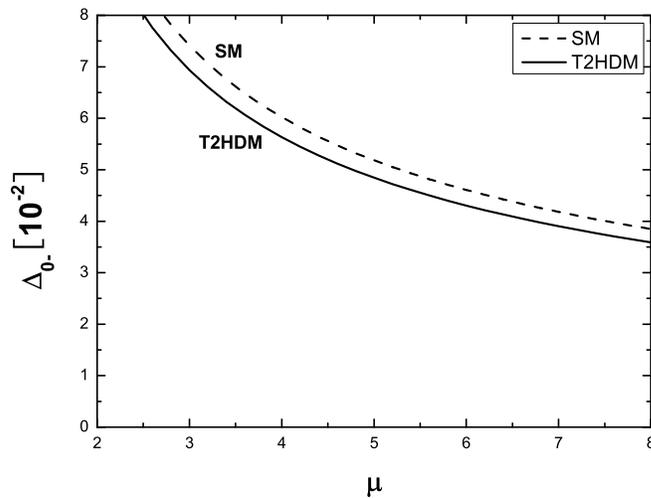}}}
\vspace{-1cm}
\caption{ The $\mu$ dependence of the isospin
symmetry breaking $\Delta_{0-}(K^*\gamma)$ in the SM and T2HDM's.
For details see the text.} \label{fig:fig6}
\end{figure}

\begin{figure}[htb] 
\vspace{-1cm}
\centerline{\mbox{\epsfxsize=10cm\epsffile{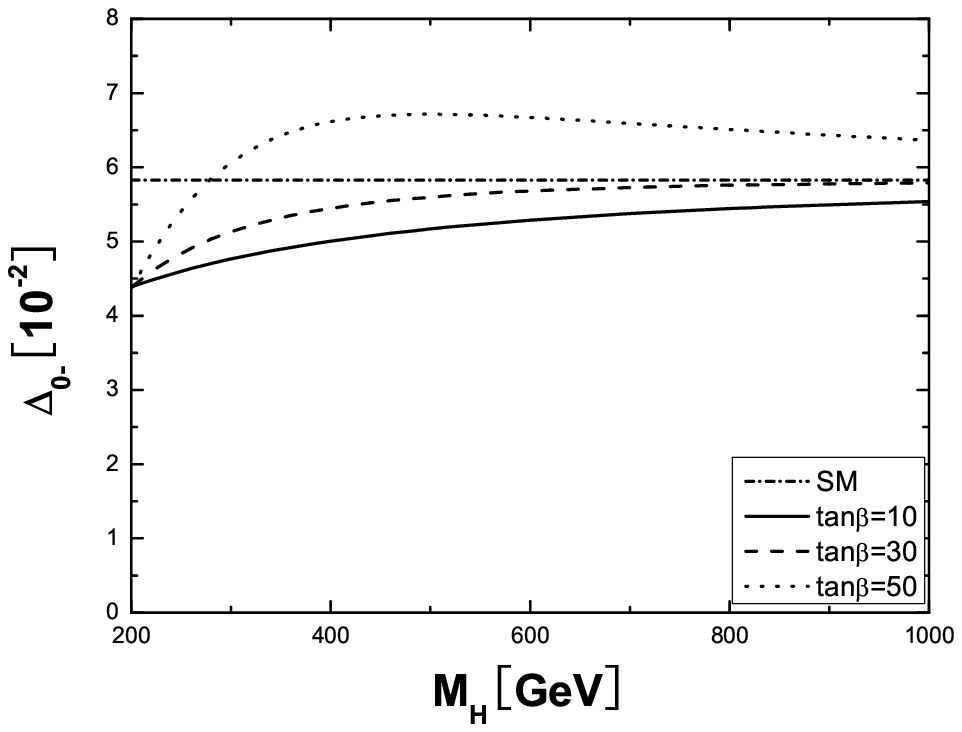}}}
\vspace{-1cm}
\caption{The $\mhp$ dependence of the isospin
symmetry breaking $\Delta_{0-}(K^*\gamma)$ in the SM and the
T2HDM. For details see the text.} \label{fig:fig7}
\end{figure}

In the top-quark two-Higgs-doublet model, and assuming $\xi=1$, $\tan{\beta}=30$
and $\mhp=400$ GeV, we find numerically that
\beq
\Delta_{0-}(K^*\gamma)^{\rm T2HDM} &=& \left [ 5.5^{+3.8}_{-2.0} (\mu) ^{+1.5}_{-1.0}(F_{K^*})
 ^{+0.5}_{-1.1}(\lambda_B) \pm 0.3 (m_c)\right ] \times 10 ^{-2} \non
 &=&  \left ( 5.5 ^{+4.1}_{-2.5} \right )\times 10^{-2}
 \eeq
where the individual errors have been added in  quadrature, the dominant error comes from
the uncertainty of the energy scale $\mu$.

In Fig.~\ref{fig:fig6}, we show the $\mu$ dependence of the isospin
symmetry breaking $\Delta_{0-}(K^* \gamma)$ in the SM and the T2HDM for
$\xi=1$, $\tan{\beta}=30$ and $\mhp=400$ GeV. The dashed and solid curve shows the central value of
the SM  and T2HDM prediction, respectively.
One can see from this figure that the new physics effects on the isospin symmetry breaking is very small.
The reason is that the new physics contributions to the related decays are largely canceled
in the ratio.

Of course, the isospin symmetry breaking in the T2HDM also have a moderate dependence
on the values of $\mhp$, $\tan\beta$ and $\xi$.
Fig.~\ref{fig:fig7}  shows the $\mhp$ and $\tan\beta$ dependence of the isospin
symmetry breaking $\Delta_{0-}(K^* \gamma)$ in the T2HDM's for $\xi=1$.
The solid, dashed and dots curve shows the T2HDM predictions for
$\tan{\beta}=10,30$ and $50$, respectively. The dot-dash line
shows the SM prediction. The whole region of Fig.~\ref{fig:fig6} and \ref{fig:fig7} is allowed by the
data: $\Delta_{0-}(K^*\gamma)^{exp}= (3.9\pm 4.2)\times 10^{-2}$.

\section{$B \to \rho \gamma$ decay} \label{sec:brho}

When compared with $B \to K^* \gamma$ decay, the $B \to \rho
\gamma$ decay mode is particularly interesting in search for new
physics beyond the SM.
Firstly, its branching ratio will be suppressed with respect to $B \to K^* \gamma$ by roughly
a factor of  $|V_{td}/V_{ts}|^2 \approx 4 \times 10^{-2}$.
In contrast to the $B \to K^* \gamma$ decay, the CP-violating asymmetry for $B \to \rho \gamma$ decay
is generally at $10\%$ level and may be observed in B factory experiments.

\subsection{Branching ratios and CP asymmetries}

From Eq.~(\ref{eq:br-vga}), the branching ratios of $B \to \rho
\gamma$ decays can be written as
\beq
\calb (B \to \rho
\gamma)& =& \tau_B\frac{G_F^2\alpha m_B^3m_b^2}{32\pi^4} \left
(1-\frac{m_\rho^2}{m_B^2} \right )^3 \left | R_{\rho} \right |^2
c_\rho^2  |F_\rho|^2,
\label{eq:br-rho}
\eeq
with
\beq
R_{\rho} =V_{ud}^* V_{ub} \,\left [  a_7^u(\rho \gamma) + a_{ann}^u (\rho
\gamma) \right ] +  V_{cd}^* V_{cb}\,\left [  a_7^c(\rho \gamma)
+a_{ann}^c (\rho \gamma) \right ].
\label{eq:rrho}
\eeq
Using the input parameters as given in Table \ref{input}, we find the
SM predictions for the branching ratios and CP-violating asymmetries of  $B \to \rho \gamma$
decays
\beq
\calb (\overline{B}^0 \to \rho^0 \gamma )^{\smallsm} &=&
\left [ 0.77  ^{+0.21}_{-0.18}(F_{\rho}) ^{+0.09}_{-0.12} (\mu_b) ^{+0.07}_{-0.02}(\lambda_B)
^{+0.19}_{-0.15} (\gamma) \right ] \times 10 ^{-6} \non
&=& \left ( 0.77 ^{+0.35}_{-0.30}  \right ) \times 10^{-6},
\label{eq:br-smr1}\\
\calb (B^- \to \rho^- \gamma )^{\smallsm} &=&\left [ 1.66 \pm 0.4 (F_\rho)
^{+0.07}_{-0.11}(\mu_b) \pm 0.3 (\lambda_B) ^{+0.20}_{-0.16}(\gamma) \right ] \times
10 ^{-6} \non &=& \left ( 1.66 \pm 0.58  \right ) \times
10^{-6}, \label{eq:br-smr2}\\
\acp(\rho^0 \gamma )^{\smallsm} &=& \left [  8.3 ^{+3.8}_{-1.8} (\mu_b)^{+1.5}_{-1.6} (R_b\&\gamma)
^{+0.8}_{-1.6} (\lambda_B) ^{+0.9}_{-1.1} (m_c) \right ] \times 10^{-2}\non
&=& \left ( 8.3 ^{+4.3}_{-3.2}  \right ) \times 10^{-2}, \label{eq:acp-sm1}\\
\acp (\rho^\pm \gamma )^{\smallsm} &=& \left [  10.3 ^{+5.4}_{-2.5}
(\mu_b)^{+1.8}_{-2.0} (R_b\&\gamma) \pm 0.8 (m_c) \pm {0.1} (\lambda_B) \right ] \times
10^{-2}\non &=& \left (  10.3^{+5.8}_{-3.5}  \right ) \times
10^{-2}, \label{eq:acp-sm2}
\eeq
where the individual errors coming from uncertainties of $F_\rho$, $\mu_b$, $\lambda_B$, $m_c$, $R_b$ and
$\gamma$ have been taken into account and added in quadrature.
As expected, the dominant error is induced by the uncertainty of the form factor $F_\rho$
for the branching ratios,
and by the uncertainty of the scale $\mu_b$ and the parameter $R_b$ and $\gamma$ for the CP-violating
asymmetries.

For $\mhp =400$ GeV, $\tan\beta=30$ and $\xi=1$, the theoretical predictions for the
branching ratios and CP-violating asymmetries of $B \to \rho \gamma$ decays in the T2HDM are
as follows
\beq
\calb (\rho^0 \gamma )^{\smallnp} &=&
\left ( 0.88 ^{+0.37}_{-0.34}  \right ) \times 10^{-6},
\label{eq:br-np1}\\
\calb (\rho^- \gamma )^{\smallnp} &=& \left ( 1.86 ^{+0.74}_{0.63}  \right ) \times
10^{-6}, \label{eq:br-np2}\\
\acp(\rho^0 \gamma )^{\smallnp} &=&
\left ( 7.8 ^{+3.9}_{-2.7}  \right ) \times 10^{-2}, \label{eq:acp-np1}\\
\acp (\rho^\pm \gamma )^{\smallnp} &=& \left (  11.1^{+6.0}_{-3.7}  \right ) \times
10^{-2}, \label{eq:acp-np2}
\eeq
where the individual errors coming from uncertainties of $F_\rho$, $\mu_b$, $\lambda_B$, $m_c$, $R_b$ and
$\gamma$ have been taken into account and added in quadrature.

\begin{figure}[htb]  
\vspace{-1cm}
\centerline{\mbox{\epsfxsize=10cm\epsffile{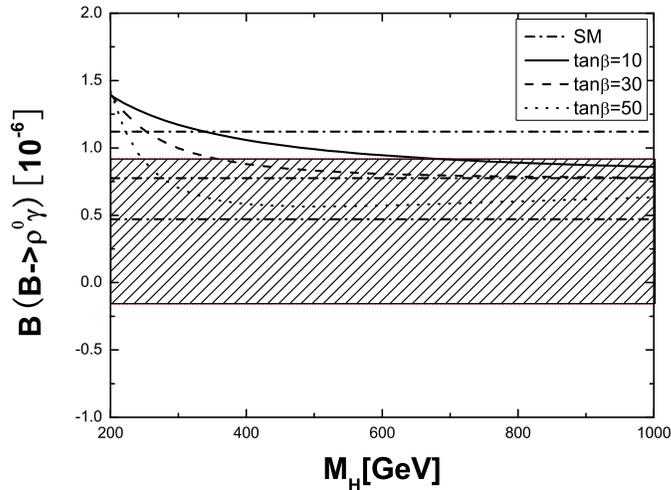}}}
\vspace{-1cm} \caption{Plots of the $\mhp$ dependence of the
branching ratio $\calb(B \to \rho^0 \gamma)$ in the SM (short-dash
curves) and T2HDM (solid, dashed and dots curve for
$\tan\beta=10,30,50$,respectively). The shaded band shows the data
at $3\sigma$ level.} \label{fig:fig8}
\end{figure}

\begin{figure}[htb] 
\vspace{-1cm}
\centerline{\mbox{\epsfxsize=10cm\epsffile{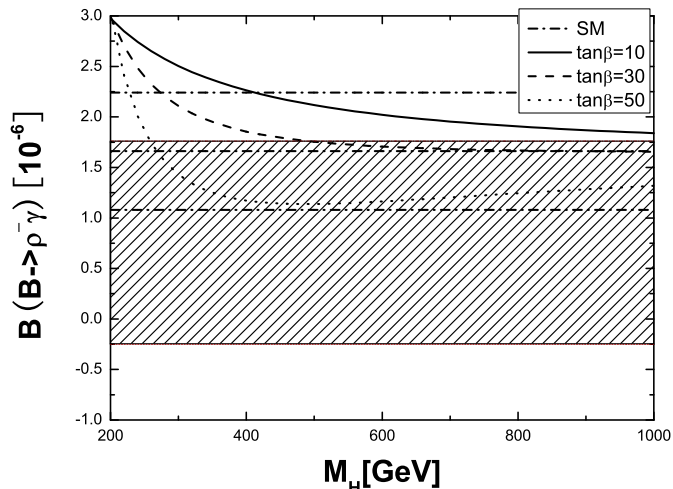}}}
\vspace{-1cm} \caption{The same as Fig.~\ref{fig:fig8} but for $B
\to \rho^- \gamma$ decay. } \label{fig:fig9}
\end{figure}

In Fig.~\ref{fig:fig8}, we show the $\mhp$ dependence of the branching ratio
$\calb (B \to \rho^0 \gamma)$ in the T2HDM for
$\xi=1$, $\tan{\beta}=10$(solid curve), $30$ (dashed curve) and $50$ GeV (dotted curve),
respectively. The central value and the error of the SM prediction as given in Eq.(\ref{eq:br-sm1})
are also shown by three horizontal dot-dash lines.
Fig.~\ref{fig:fig9} shows the $\mhp$ dependence of the branching ratio as in Fig.~\ref{fig:fig8}
but for $B^- \to \rho^- \gamma$ decay.

\begin{figure}[htb] 
\vspace{-1cm}
\centerline{\mbox{\epsfxsize=10cm\epsffile{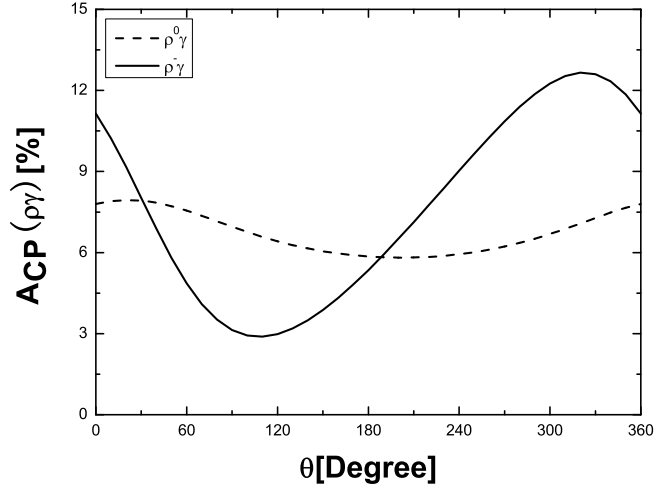}}}
\vspace{-1cm}
\caption{Plots of $\theta$ dependence of the CP-violating asymmetries for
$B \to \rho^0 \gamma$ (dashed curve) and $B^\pm \to \rho^\pm \gamma$ (solid curve) decay
in the T2HDM, assuming $\mhp=400$ GeV, $\tan\beta=30$ and $|\xi|=1$. }
\label{fig:fig10}
\end{figure}

\begin{figure}[tb]  
\vspace{-1cm}
\centerline{\mbox{\epsfxsize=10cm\epsffile{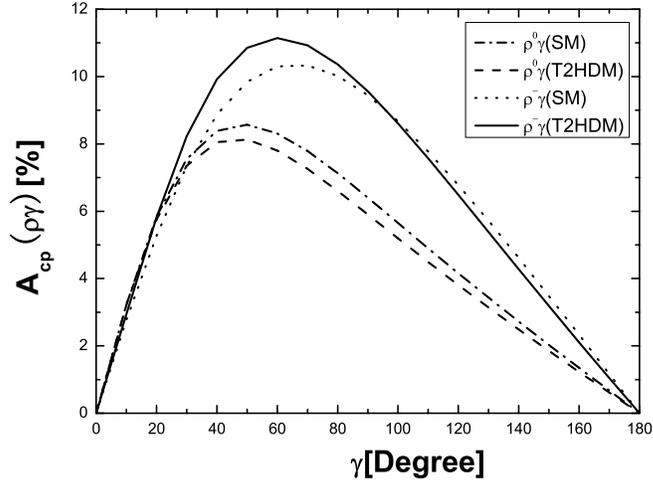}}}
\vspace{-1cm}
\caption{Plots of the CKM angle $\gamma$ dependence of
the CP asymmetries for  $B \to \rho \gamma$ decays in the SM
and T2HDM, assuming $\mhp=400$ GeV, $\tan\beta=30$ and $\xi=1$.}
\label{fig:fig11}
\end{figure}

In the T2HDM, the CP-violating asymmetry has only a weak dependence on the value of
$\mhp$: $6.3\% \leq \acp(B \to \rho^0\gamma) \leq 8.3\%$ and
$7.7\% \leq \acp(B \to \rho^\pm \gamma) \leq 11.1\%$ for $300 {\rm GeV} \leq \mhp \leq 700$ GeV.
But it has a strong dependence on the angle $\theta$ as expected.
Fig.~\ref{fig:fig10} shows the $\theta$ dependence of the CP-violating asymmetry
for $B \to \rho^0 \gamma$ ( dashed curve) and $\rho^\pm \gamma$ decay (solid curve),
assuming $\tan\beta=30$, $|\xi|=1$ and $\mhp=400$ GeV.
Fig.~\ref{fig:fig11} shows the CKM angle $\gamma$ dependence of the CP-violating asymmetry
for $B \to \rho^0 \gamma$ ( dashed curve) and $\rho^\pm \gamma$ decay (solid curve),
assuming $\tan\beta=30$, $\xi=1$ and $\mhp=400$ GeV. The SM predictions are also shown
by dot-dash ($\rho^0\gamma$ channel ) and dotted curve ($\rho^\pm \gamma$ channel),
respectively.

\newpage

By comparing the above theoretical predictions in the SM and T2HDM with the measured values of
branching ratios as given in Eqs.(\ref{eq:brho-exp}) and (\ref{eq:bro-exp}), we find that
\begin{enumerate}
\item[]{(i)}
 The central values of the theoretical predictions for the branching ratios are generally
larger than the measured
values, but still consistent with them within two standard deviation, as shown in Figs.~\ref{fig:fig8}
and \ref{fig:fig9}. Since the theoretical predictions for the branching ratios are proportional to
the value of the form factor $F_\rho$, the current data clearly prefer a smaller $F_\rho$, just like the
case of $B \to K^* \gamma$ decays.

\item[]{(ii)}
The new physics contributions tend to increase the branching ratios for light charged-Higgs boson,
but such enhancement become smaller rapidly when the charged-Higgs boson becoming heavy, as illustrated
in Figs.~\ref{fig:fig8} and \ref{fig:fig9}. For a charged-Higgs boson with a mass around $400$ GeV,
the theoretical predictions in the T2HDM become compatible with the data and the SM predictions.

\item[]{(iii)}
In the T2HDM, the CP-violating asymmetry has a strong $\theta$ dependence for $B^\pm \to \rho^\pm \gamma$
decay: $2.9\% \leq \acp(B \to \rho^\pm \gamma) \leq 12.7\%$.

\item[]{(iv)}
 One can see from the numerical results that a charged-Higgs boson with a mass of $300-500$ GeV
is still allowed by the data of $B \to \rho \gamma$.

\end{enumerate}

\subsection{Isospin and U-spin symmetries }

In Refs.~\cite{bosch02a,bosch02b}, the isospin symmetry breaking of $B \to \rho \gamma$ decays
has been defined as the form of
\beq
\Delta (\rho\gamma) = \frac{1}{2}\left [\frac{\Gamma(B^+ \to \rho^+ \gamma)}{2\Gamma(B^0\to \rho^0
\gamma)} + \frac{\Gamma(B^- \to \rho^- \gamma)}{
2\Gamma(\bar{B^0}\to \rho^0 \gamma)} -2 \right ].
\label{eq:iso1}
\eeq

Using the central values of input parameters as listed in Table
\ref{input} and assuming $tan{\beta}=30$, $\mhp=400$ GeV, we find
numerically
\beq
\Delta (\rho\gamma)&=& \left \{\begin{array}{ll}
\left ( 0.7 ^{+19.0}_{-11.5} \right )\times 10^{-2}, & {\rm in \ \ SM}, \\
\left ( 0.4 ^{+17.6}_{-10.2} \right )\times 10^{-2}, & {\rm in \ \ T2HDM},
\end{array} \right.
\label{eq:iso2}
\eeq
where the errors coming from uncertainties of input parameters have
been added in quadrature. The uncertainty of the CKM angle
$\gamma$ and the parameter $\lambda_B$ dominate the total theoretical error.
Although the central value of the isospin breaking $\Delta (\rho\gamma)$ is very small, but it can be as large
as $10\% - 20\%$ in the parameter space considered. The new physics correction on
$\Delta (\rho\gamma)$ is too small to be separated experimentally.

Under the approximation $\Gamma(B^+ \to \rho^+ \gamma) = \Gamma(B^- \to \rho^- \gamma)$, and
$\Gamma(B^0 \to \rho^0 \gamma) = \Gamma(\overline{B}^0 \to \rho^0 \gamma)$, Eq.~(\ref{eq:iso1}) can be
rewritten as
\beq
\Delta (\rho\gamma)^{exp} = \frac{1}{2}\left [\frac{\tau_{B^0}}{\tau_{B^+}}
\frac{\calb (B \to \rho^\pm \gamma)^{exp}}{\calb (B \to \rho^0 \gamma)^{exp}} -2 \right ],
\label{eq:iso3}
\eeq
and we find numerically
\beq
\Delta (\rho\gamma)^{exp} = -0.18^{+0.44}_{-0.38}(\calb (\rho^\pm \gamma))
^{+0.74}_{-0.26}(\calb (\rho^0 \gamma))
= -0.18 ^{+0.86}_{-0.46}
\label{eq:iso4}
\eeq
by using the measured values of the branching ratios and lifetimes
as given in Eq.(\ref{eq:bro-exp}) and in Ref.~\cite{pdg2004}.
Although the central value
of $\Delta (\rho\gamma)^{exp}$ is in the reasonable region, but its error
is too large to compare meaningfully with the theoretical predictions.

\begin{figure}[tb]  
\vspace{-1cm}
\centerline{\mbox{\epsfxsize=10cm\epsffile{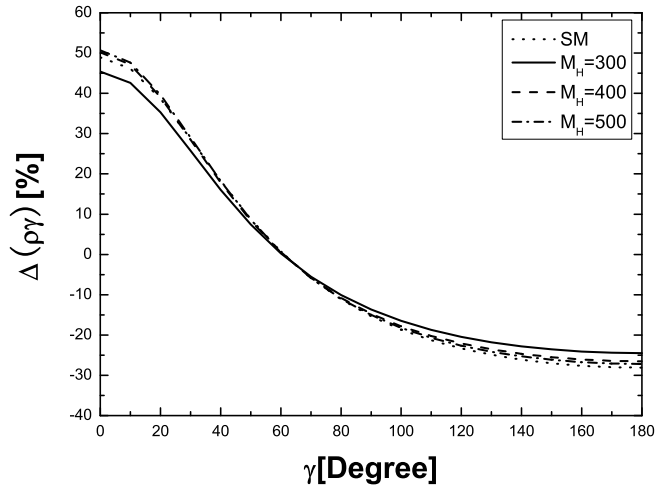}}}
\vspace{-1cm}
\caption{The isospin breaking $\Delta(\rho \gamma)$
vs the CKM angle $\gamma$ in the SM (dots curves) and
T2HDM (solid curve).}
\label{fig:fig12}
\end{figure}

In \fig{fig:fig12}, we show the angle $\gamma$ dependence of the
isospin symmetry breaking $\Delta (\rho \gamma)$ in the SM and the T2HDM
for $\tan{\beta}=30$, $\xi=1$ and $\mhp=300$ (solid curve), $500$ (dashed curve) and
$700$ GeV (dotted curve). It is easy to  see from \fig{fig:fig12}
that
\begin{itemize}
\item[]{(i)}
The isospin breaking in the SM and T2HDM's have the
similar $\gamma$ dependence;

\item[]{(ii)}
All theoretical predictions become almost identical and very small in magnitude for
$\gamma \sim 60^\circ$, the value of $\gamma$ preferred by the global fit.
The smallness of $\Delta (\rho\gamma)$ is also consistent with the
general expectation and other measurements;

\item[]{(iii)}
The theoretical predictions in the SM and T2HDM have the similar $\gamma$
dependence, and have the same sign for small or large values of
the CKM angle $\gamma$.

\end{itemize}

Another interesting observable for $B \to (K^*, \rho) \gamma$
decays is the U-spin symmetry, it has been studied in
Refs.\cite{bosch02a,bosch02b,isosm}. In the limit of U-spin
symmetry, the quantity
\beq
\Delta U(K^*,\rho)\equiv \Delta
\calb(B \to K^* \gamma) + \Delta \calb(B \to \rho \gamma) \equiv 0
\eeq with \beq \Delta \calb(B \to K^* \gamma) &=& \calb (B^+ \to
K^{*+} \gamma) - \calb (B^- \to K^{*-}
\gamma), \\
\Delta \calb(B \to \rho \gamma)&=& \calb (B^+ \to \rho^+ \gamma) -
\calb (B^- \to \rho^- \gamma),
\eeq
should be satisfied. Using the
central values of input parameters, we find the SM predictions for
$\Delta \calb(B \to K^* \gamma) $ and $\Delta \calb(B \to K^* \gamma) $
\beq
\Delta \calb(B \to K^* \gamma) &=&-3.3\times 10^{-7}, \quad
\Delta \calb(B \to \rho \gamma) = +3.8\times 10^{-7}.
\eeq
Here we have chosen $\gamma=60^\circ$ which maximizes the effects. The two
parts have opposite sign and cancels to a large extent, leaving a
small U-spin breaking
\beq
\Delta U(K^*, \rho) =0.5\times 10^{-7}\label{eq:dsb1}.
\eeq
in the SM, which is only about $8\%$ of the branching ratio $\calb (B \to \rho^0 \gamma)$.
In the T2HDM, we find the numerical result
\beq
\Delta U(K^*, \rho)&=& 1.1  \times 10^{-7}, \label{eq:usb1}
\eeq
for $\tan{\beta}=30$, $\mhp=400$ GeV and $\gamma=60^\circ$.
The new physics contribution in the conventional T2HDM has little
effect on the size of U-spin symmetry breaking.

\section{Conclusions}

By employing the QCD factorization approach for the exclusive $B
\to V \gamma$ decays as proposed in
Refs.\cite{beneke01,ali02a,bosch02a}, we calculated the new
physics contributions to the branching ratios, CP asymmetries,
isospin symmetry breaking and U-spin symmetry breaking of the
exclusive radiative decays $B \to K^*\gamma$ and $B \to \rho
\gamma$, induced by the charged Higgs penguin diagrams appeared in
the top-quark two-Higgs-doublet model \cite{das96,wu99,kiers00}.
The new physics
contributions are included through their  corrections to the Wilson
coefficients $C_7(\mw)$ and $C_8(\mw)$ at the matching
scale $\mw$.

In section \ref{sec:th},we describe briefly the basic structures
of the T2HDM, give a brief review
about the calculation of $B \to V \gamma$ ($V=K^*, \rho$) at NLO
in QCD factorization approach and present the needed analytical
formulas.

In section \ref{sec:bks} and \ref{sec:brho}, we calculated the new
physics contributions to the physical observable of $B \to K^* \gamma$ and $B
\to \rho \gamma$ decays in the T2HDM, compared the
theoretical predictions with those currently available
experimental measurements, and we found that:
\begin{itemize}

\item[]{(i)}
In the T2HDM, a light charged-Higgs boson with a mass less than $200$ GeV is clearly
excluded by the date of $B \to V \gamma$ decay, but a charged-Higgs boson
with a mass larger than $300$ GeV are always allowed by the same set of data.
Such lower limits on $\mhp$ are comparable with those obtained from the
inclusive $B \to X_s \gamma$ decay.

\item[]{(ii)}
In the SM and T2HDM, the theoretical predictions for CP
asymmetry of $B \to K^* \gamma$ is always less than $1\%$ in size,
but CP asymmetry of $B \to \rho \gamma$ can be as large as $10\%$ in
magnitude and have a strong dependence on the variations of the
angle $\theta$, the scale $\mu={\cal O}(m_b)$ and the CKM angle $\gamma$.

\item[]{(iii)}
The isospin symmetry breaking for $B \to V\gamma$ decays
in the SM and T2HDM considered here are generally small in size:
around $6\%$ for $B \to K^{*} \gamma$ decay (well consistent with the data ),
and in the range of $[-0.11, 0.20]$
for $B \to \rho \gamma$ when the effects of theoretical uncertainties of input parameters
are also taken into account.

\item[]{(iv)}
The U-spin symmetry breaking $\Delta U(K^*,\rho)$ in the SM
and T2HDM's considered here is generally small in size, only about $8\%$ of the
branching ratio $\calb (B \to \rho^0 \gamma)$.

\end{itemize}

\begin{acknowledgments}

We are very grateful to Li-bo Guo for helpful discussions. This
work is partially supported  by the National Natural Science
Foundation of China under Grant No.10275035, 10575052 and by the
Specialized Research Fund for the doctoral Program of higher education (SRFDP)
under Grant No.~20050319008.

\end{acknowledgments}


\begin{appendix}

\section{Wilson coefficients}\label{app:wcs}

At the low energy scale $\mu =O(m_b)$, the leading order Wilson
coefficients  are
\beq
 C_{j,\smallsm}^{0}(\mu) &=& \sum_{i=1}^{8}k_{ji}\eta^{a_i},  \ \  for \ \ j = 1,...,6,\\
 C_{7,\smallsm}^{0}(\mu) &=&
 \eta^{\frac{16}{23}}C_{7,\smallsm}^{0}(M_{W})
 +\frac{8}{3}(\eta^{\frac{14}{23}}-\eta^{\frac{16}{23}})C_{8,\smallsm}^{0}(M_W)
 +\sum_{i=1}^{8}h_i\eta^{a_i}, \label{eq:c70mb}\\
 C_{8,\smallsm}^{0}(\mu) &=& C_{8,\smallsm}^{0}(M_W)\eta^{\frac{14}{23}}
 + \sum_{i=1}^{8}\hbar_i\eta^{a_i}
\label{eq:c80mb}
 \eeq
in the standard basis, while \beq
 Z_{j,\smallsm}^{0}(\mu) &=& \sum_{i=1}^{8}{h_{ji}\eta^{a_i}}, \ \  for\ \ j = 1...6,\\
 Z_{7,\smallsm}^{0}(\mu) &=& C_{7,\smallsm}^{0}(\mu),\label{eq:z70mb}\\
 Z_{8,\smallsm}^{0}(\mu) &=& C_{8,\smallsm}^{0}(\mu)\label{eq;z80mb}
 \eeq
in the CMM basis. Here $\eta=\alpha_s(\mw)/\alpha_s(\mub)$, and the expressions of the
Wilson coefficients
$C_{7,\smallsm}^{0}(\mw)$ and $C_{8,\smallsm}^{0}(\mw)$, the``magic numbers"
$a_i$, $k_{ji}$, $h_{ji}$, $h_i$ and $\hbar_i$ can be found in Ref.~~\cite{buras96}.

The NLO Wilson coefficient $C_7(\mub)$ at scale $\mub={\cal O}(m_b)$ can be written as
\beq
C_{7,\smallsm}(\mu) = C_{7,\smallsm}^0(\mu) + \frac{\alpha_s(\mu)}{4\pi}
C_{7,\smallsm}^1(\mu)\label{eq:c7nlo}
 \eeq
with
\beq
C_{7,\smallsm}^{1}(\mu) &=&
\eta^{\frac{39}{23}}C_{7,\smallsm}^{1}(M_{W})+\frac{8}{3}
                 \left (\eta^{\frac{37}{23}}-\eta^{\frac{39}{23}} \right )C_{8,\smallsm}^{1}(M_W)\non
            &&  +\left ( \frac{297664}{14283}\eta^{\frac{16}{23}}
                -\frac{7164416}{357075}\eta^{\frac{14}{23}}
                +\frac{256868}{14283}\eta^{\frac{37}{23}}
                -\frac{6698884}{357075}\eta^{\frac{39}{23}}\right ) C_{8,\smallsm}^{0}(M_W)\non
            &&  +\frac{37208}{4761}\left ( \eta^{\frac{39}{23}}-\eta^{\frac{16}{23}} \right )
            C_{7,\smallsm}^{0}(M_W)
                +\sum_{i=1}^{8}{(e_i\eta E(x_t)+ f_i + g_i\eta)\eta^{a_i}},
\eeq
where the function $E(x_t)$ and the ``magic numbers" $e_i$, $f_i$ and $g_i$
can also be found in Ref.~\cite{buras96}.

\section{$G_i$ and $H_i^V$ functions}\label{app:gi}

In this Appendix, the explicit expressions or numerical values of
all $G_i$ and $H_i^V$ functions appeared in Eq.~(\ref{eq:avga})
will be listed. For more details of these functions, one can see
Ref.~\cite{bosch02b} and references therein.

\beq
G_1(z) &=& \displaystyle \frac{52}{81}\ln\frac{\mu}{m_b}
    +\frac{833}{972}-\frac{1}{4}[a(z)+b(z)]+\frac{10i\pi}{81}, \\
G_2(z) &=& \displaystyle -\frac{104}{27}\ln\frac{\mu}{m_b}
     -\frac{833}{162} +\frac{3}{2}[a(z) +b(z)]-\frac{20i\pi}{27}, \\
G_3      &=& \displaystyle \frac{44}{27}\ln\frac{\mu}{m_b}
    + \frac{598}{81} +\frac{2\pi}{\sqrt{3}} +\frac{8}{3} X_b
    -\frac{3}{4}a(1) +\frac{3}{2}b(1) +\frac{14i\pi}{27}, \\
G_4(z_c) &=& \displaystyle \frac{38}{81}\ln\frac{\mu}{m_b}
    + -\frac{761}{972} -\frac{\pi}{3\sqrt{3}} -\frac{4}{9} X_b
    +\frac{1}{8} a(1) +\frac{5}{4} b(z_c) -\frac{37i\pi}{81} \\
G_5      &=& \displaystyle \frac{1568}{27}\ln\frac{\mu}{m_b}
    + \frac{14170}{81} +\frac{8\pi}{\sqrt{3}} +\frac{32}{3} X_b
    -12 a(1) +24 b(1) +\frac{224i\pi}{27},       \\
G_6(z_c) &=& \displaystyle -\frac{1156}{81}\ln\frac{\mu}{m_b}
    + \frac{2855}{486} -\frac{4\pi}{3\sqrt{3}} -\frac{16}{9} X_b\non
    && -\frac{5}{2} a(1) +11 b(1) +9 a(z_c) +15 b(z_c) -\frac{574i\pi}{81}, \\
G_8      &=& \displaystyle \frac{8}{3}\ln\frac{\mu}{m_b}
    + \frac{11}{3} -\frac{2\pi^2}{9} +\frac{2i\pi}{3}, \\
\eeq
where
\beq
X_b &=& \int_0^1 \!dx \int_0^1 \!dy \int_0^1 \! dv x y \ln[v+x(1-x)(1-v)(1-v+v y)]
\approx -0.1684, \label{eq:xb}\\
a(1) &\simeq & 4.0859 + \frac{4i\pi}{9},\label{eq:a1} \\
b(1) &=& \frac{320}{81} - \frac{4 \pi}{3 \sqrt{3}} +
\frac{632\pi^2}{1215}
  - \frac{8}{45} \left[ \frac{d^2 \ln \Gamma(x)}{dx^2} \right]_{x=\frac{1}{6}}+ \frac{4i\pi}{81}
  \simeq 0.0316 + \frac{4i\pi}{81}, \label{eq:b1}\\
a(z_u)&=& \left ( -1.93 + 4.96 i \right )\times 10^{-5}, \label{eq:azu}\\
a(z_c)&=&   1.525 + 1.242 i , \label{eq:azc}\\
b(z_u)&=& \left ( 1.11 + 0.28 i \right )\times 10^{-5}, \label{eq:bzu}\\
b(z_c)&=&  -0.0195 + 0.1318 i , \label{eq:bzc}
\eeq
where $z_q=m_q^2/m_b^2$ and the masses $m_q$ ($q=u,c,b$) as listed in
\tab{input} have been used to obtain the numerical results. The
explicit analytical expressions for $a(z)$ and $b(z)$ can be found
for example in Ref.~\cite{bosch02b}.

For the $H_i^V$ functions, we have
\beq
H^V_1(z_p)   &=& -\frac{2\pi^2}{9}\frac{f_B f^\perp_V }{F_V m_B\lambda_B}
  \int^1_0 dv\, h(\bar v,z_p) \Phi_V^\perp(v),   \label{eq:h1v}\\
H^V_2      &=& 0\\
H^V_3      &=& -\frac{1}{2}\left[ H^V_1(1) +H^V_1(0)\right], \\
H^V_4(z_c) &=&  H^V_1(z_c)-\frac{1}{2}H^V_1(1), \label{eq:h4v}\\
H^V_5      &=&  2 H^V_1(1), \label{eq:h5v}\\
H^V_6(z_c) &=& -H^V_1(z_c)+\frac{1}{2}H^V_1(1) =-H^V_4(z_c), \label{eq:h6v}\\
H^V_8   &=& +\frac{4\pi^2}{3}\frac{f_B f^\perp_V }{F_V
m_B\lambda_B}
  \left ( 1 -\alpha_1^V + \alpha_2^V + \cdots \right ),   \label{eq:h8v}
\eeq
where the hard-scattering function $h(u,z)$ is given by
\beq
h(u,z)&=& \frac{4z}{u^2}\left\{  Li_2\!\left[
\frac{2}{1-\sqrt{\frac{u-4z+i\varepsilon}{u}}}\right ] +
Li_2\!\left[ \frac{2}{1+\sqrt{\frac{u-4z+i\varepsilon}{u}}}\right
]\right \} -\frac{2}{u},
\eeq
where $Li_2[x]$ is the dilogarithmic function, and the function $h(u,z)$ is real for $u \leq 4z$ and
develops an imaginary part for $u > 4z$.  The light-cone wave
function $\Phi_V^\perp(v)$ takes the form of
\beq
\Phi_V^\perp (v)&=& 6 v (1-v) \left [ 1 +  \alpha_1^V(\mu) C_1^{3/2}(2v-1) +
\alpha_2^V(\mu) C_2^{3/2}(2v-1) + \cdots \right ]
\eeq
where $C_1^{3/2}(x)=3x$, $C_2^{3/2}(x)=\frac{3}{2}(5x^2 -1)$.

\end{appendix}


\newpage
\begin{thebibliography}{99}


\bibitem{buras96}
G.~Buchalla, A.J.~Buras, and M.E.~Lautenbacher, \rmp {\bf 68},
1125 (1996).

\bibitem{hurth02}
T.~Hurth, \rmp {\bf 75}, 1159(2003).

\bibitem{hbb05}
T.~Hurth, talk presented at Beauty 2005, Perugia, Italy, 20 -24 June, 2005;
S.W.~Bosch, talk presented at CKM2005 Workshop, San Diego, USA, 15-18 March, 2005;
C.~Bobeth, talk presented at CKM2005 Workshop, San Diego, USA, 15-18 March, 2005.

\bibitem{hfag05}
Heavy Flavor Averaging Group, http://www.slac.stanford.edu/xorg/hfag.

\bibitem{kn99}
A.L.~Kagan and M.~Neubert, \epjc {\bf 7}, 5 (1999);
A.J.~Buras, A.~Czarnecki, M.~Misiak, and J.~Urban, \npb {\bf 611}, 488 (2001),  {\bf 631}, 219 (2002).

\bibitem{carena01}
M.~Carena, D.~Garcia, U.~Nierste and C.~E.~Wagner, \plb {\bf 499},141 (2001);
G.~Degrassi, P.~Gambino and G.~F.~Giudice, JHEP {\bf 0012}, 009 (2000);
G.~D'Ambrosio, G.~F.~Giudice, G.~Isidori and A.~Strumia, \npb {\bf 645}, 255 (2002).

\bibitem{bor00}
F.~Borzumati, C.~Greub, T.~Hurth and D.~Wyler, \prd {\bf 62}, 075005 (2000);
T.~Besmer, C.~Greub and T.~Hurth, \npb {\bf 609}, 359 (2001).

\bibitem{2hdm}
S.~Glashow and S.~Weinberg, \prd {\bf 15}, 1958 (1977);
J.F.~Gunion, H.E.~Haber, G.~Kane, and S.~Dawson, {\em The Higgs
Hunter's Guide}, Addison Wesley, Redwood-City (1990), and
references therein.

\bibitem{xiao04}
Z.J.~Xiao and L.B.~Guo, \prd 69, 014002 (2004).

\bibitem{belle-bd}   
D.~Mohapatra {\it et al.}, Belle Collab., \prd 72, 011101(R) (2005);
K.~Abe {\it et al.}, Belle Collab., hep-ex/0506079.

\bibitem{deshpande87}
N.G.~Deshpande, P.~Lo, J.~Trampetic, G.~Eilam, and P.~Singer, \prl {\bf 59}, 183 (1987).

\bibitem{greub95}
C.~Greub, H.~Simma and D.~Wyler, \npb {\bf 434}, 39 (1995);
H.H.~Asatryan, H.M.~Asatrian, and D.~Wyler, \plb {\bf 470}, 223 (1999).

\bibitem{beneke01}
M.~Beneke, T.~Feldmann and D.~Seidel, \npb {\bf 612}, 25 (2001).

\bibitem{ali02a}
A.~Ali and A.Y.~Parkhomenko, \epjc {\bf 23}, 89 (2002).

\bibitem{bosch02a}
S.W.~Bosch and G.~Buchalla, \npb {\bf 621}, 459 (2002).


\bibitem{bosch02b}
S.W.~Bosch,  { \em Exclusive Radiative Decays of B Mesons in QCD
Factorization}, PH.D thesis,  hep-ph/0208203.

\bibitem{kagan02}
A.L.~Kagan and M.~Neubert, \plb  {\bf 539}, 227 (2002).

\bibitem{li99}
H.-n.~Li and G.L.~Lin, \prd {\bf 60}, 054001 (1999).

\bibitem{lmsy05}
C.D.~L\"u, M.~Matsumori, A.I.~Sanda and M.Z.~Yang, \prd {\bf 72}, 094005 (2005).


\bibitem{ali01}  
A.~Ali, T.~Handoko, and D.~London, \prd {\bf 63}, 014014 (2001);
A.Arhrib, C.K.~Chua and W.S.~Hou, \epjc {\bf 21}, 567 (2001);
A.~Ali and E.~Lunghi, \epjc {\bf 26}, 195 (2002).

\bibitem{xz04}
Z.J.~Xiao and C.~Zhuang, \epjc {\bf 33}, 349 (2004).


\bibitem{das96}
A.Das and C.Kao, \plb {\bf 372}, 106 (1996).

\bibitem{wu99}
K.~Kiers, A.~Soni, and G.H.~Wu, \prd {\bf 59}, 096001 (1999);
G.H.~Wu and A.~Soni, \prd {\bf 62}, 056005 (2000).

\bibitem{kiers00}
K.~Kiers, A.~Soni, and G.H.~Wu, \prd {\bf 62}, 116004 (2000).

\bibitem{ckm}  
M.~Kobayashi, T.~Maskawa, Prog. Theor. Phys. {\bf 49}, 652 (1973).

\bibitem{cmm97}
K.G.~Chetyrkin, M.~Misiak, M.~Munz, \plb {\bf 400}, 206(1997),
{\bf 425}, 414(E) (1998).

\bibitem{buras94}
A.J.~Buras, M.~Misiak, M.~M\"unz, and S.~Pokorski, \npb {\bf 424}, 374 (1994).

\bibitem{m3a}
T.P.~Cheng and M.~Sher, \prd {\bf 35}, 3484 (1987);
M.~Sher and Y.~Yuan, \prd {\bf 44}, 1461 (1991);
W.S.~Hou, \plb {\bf 296}, 179 (1992);
A.~Antaramian, L.J.~Hall,  and A.~Rasin, \prl {\bf 69}, 1871 (1992);
L.J.~Hall and S.~Winberg, \prd {\bf 48}, R979 (1993);
D.~Chang, W.S.~Hou,  and W.Y.~Keung, \prd {\bf 48}, 217 (1993);
Y.L.~Wu and L.~ Wolfenstein, \prl {\bf 73}, 1762 (1994);
D.~Atwood, L.~Reina and A.~Soni, \prl {\bf 75}, 3800 (1995).

\bibitem{m3b}
D.~Atwood, L.~Reina and A.~Soni, \prd {\bf 55}, 3156 (1997);
T.M.~Aliev and E.O.Iltan, \jpg {\bf 25}, 989 (1999);
D.B.~ Chao, K.~Heung, and W.Y.~ Keung, \prd {\bf 59}, 115006 (1999).

\bibitem{m3x} 
Z,J.~Xiao, C.S.~Li, and K.T.~Chao, \plb {\bf 473}, 148 (2000);
Z,J.~Xiao, C.S.~Li, and K.T.~Chao, \prd {\bf 63}, 074005 (2001);
J.J.~Cao, Z.J.~Xiao,  and G.R.~Lu, \prd {\bf 64}, 014012 (2001);
D.~Zhang, Z.J.~Xiao,  and C.S.~Li, \prd {\bf 64}, 014014 (2001);
Z.J.~Xiao, K.T.~Chao,  and C.S.~Li, \prd {\bf 65}, 114021 (2002).

\bibitem{bg98}
F.M.~Borzumati and C.~Greub,\prd{\bf 58},074004 (1998); {\it ibid } {\bf 59},057501 (1999).

\bibitem{pdg2004}
Particle Data Group, S.~Eidelman {\it et al.}, \plb 592, 1(2004).

\bibitem{ball98}
P.~Ball and V.M.~Braun, \prd {\bf 58}, 094016 (1998).

\bibitem{b03}
D.~Becirevic. talk given at the Ringberg Phenomenology Workshop on heavy flavors, Ringberg Castle,
Tegernsee, Germany, May 2003.

\bibitem{isosm}
M.~Gronau and J.L.~Rosner, \plb {\bf 500}, 247 (2001);
M.~Gronau, \plb {\bf 492}, 297 (2000);
T.~Hurth and T.~Mannel, \plb {\bf 511}, 196 (2001);
R.~Fleischer, \plb {\bf 459}, 306 (1999).

\end{thebibliography}
\end{document}